\documentclass[prb,twocolumn,showpacs,superscriptaddress,floatfix, nofootinbib]{revtex4-2}
\usepackage{amsmath,amssymb,amsfonts,mathrsfs, amsbsy}
\usepackage{graphicx}
\usepackage{lipsum, babel}
\usepackage[bookmarks=true, colorlinks=true, linkcolor=blue, urlcolor=blue, citecolor=blue, bookmarks=true, hyperindex=true]{hyperref}
\usepackage{comment}
\usepackage{bm}
\usepackage[capitalise]{cleveref}
\newcommand{\be}{\begin{equation}}
\newcommand{\ee}{\end{equation}}

\newcommand{\beq}{\begin{eqnarray}}
\newcommand{\eeq}{\end{eqnarray}}

\def\H1{\widehat{H}_1}

\newcommand{\ket}[1]{\left| #1 \right>}

\begin{document}

\title{Ergodicity in a hole-doped Anderson spin insulator}

  \author{M.\,S.~Bahovadinov}
   \email{m.bahovadinov@rqc.ru}
  \affiliation{ Laboratory for Condensed Matter Physics, National Research University Higher School of Economics, Moscow 101000, Russia}
  \affiliation{Russian Quantum Center, Skolkovo, Moscow 143025, Russia}
 \author{A.\,A.~Markov}
  \email{anton.markov@ulb.be}
 \affiliation{International Solvay Institutes, 1050 Brussels, Belgium}
  \affiliation{Center for Nonlinear Phenomena and Complex Systems, Université Libre de Bruxelles (U.L.B.), B-1050 Brussels, Belgium}
  \affiliation{Russian Quantum Center, Skolkovo, Moscow 143025, Russia}
 
   \author{G.\,V.~Shlyapnikov}
  \affiliation{Russian Quantum Center, Skolkovo, Moscow 143025, Russia}
 \affiliation{Moscow Institute of Physics and Technology, Dolgoprudny, Moscow Region,  141701, Russia}
 \affiliation{Université Paris-Saclay, CNRS, LPTMS, 91405 Orsay, France}
  \affiliation{Van der Waals–Zeeman Institute, Institute of Physics, University of Amsterdam, Science Park 904, 1098 XH Amsterdam, The Netherlands}
\date{\today}

\begin{abstract}
An interacting many-body system being localized in the presence of disorder can fail to serve as its own bath. In an electronic system, the charge and spin degrees of freedom can act as thermal baths for each other. In this paper, we study the stability of an Anderson insulator in the spin sector with respect to charge doping. We consider one-dimensional two-component fermions in a lattice with large onsite repulsion ($t-J_{XX}$ model) and with a single hole in a random magnetic field. In the absence of the hole the spin sector is localized in arbitrarily weak random magnetic fields. We argue that a single hole thermalizes the spin chain in the weak disorder limit. In this regime our numerical results indicate ergodicity of the system as a whole and delocalization of the hole. At larger magnetic fields the hole eventually localizes with a finite localization length, and the Anderson spin insulator is restored.
 \end{abstract}

\maketitle

\section{Introduction}   

\begin{figure*}[t!]
    \centering
    \includegraphics[width=0.9\textwidth]{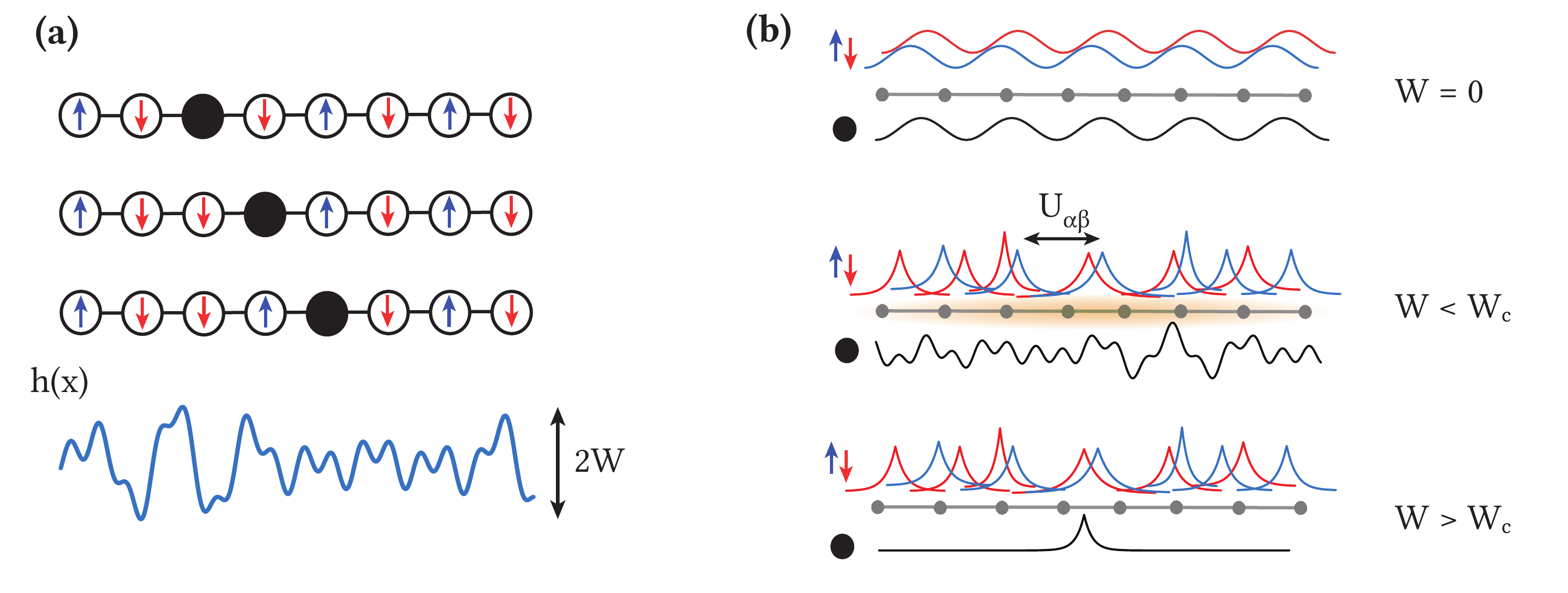}
    \caption{\textbf{Schematic illustration of the system and the main findings.} \textbf{(a)} Spin disordered $t-J_{XX}$ model doped with a single hole. \textbf{(b)} At weak disorder $W<W_c$ the spin chain is thermalized by the single hole. At a stronger disorder, $W>W_c$, the hole localizes and the system as a whole becomes an Anderson insulator.}
    \label{fig:fig1}
\end{figure*} 

The question of stability of Anderson insulator has been a major source of new ideas for decades. The very existence of Anderson insulators in nature implies~\cite{anderson1980} that the phenomenon of disorder-induced localization persists beyond the idealized Anderson model~\cite{anderson1958}. Precise conditions for the stability of the localization in a general interacting system are not yet known. It has been known from the early days of the field that the conduction within the localized band can be activated by coupling to a thermal bath via the hopping conduction mechanism~\cite{miller1960}. Ref.~\cite{basco2006} posed the long-standing problem of stability of the Anderson insulator with respect to many-body interactions (see also Refs.~\cite{anderson1980,giamarchi1987, altshuler1997, gornyi2005, abanin2019}) in an extensively studied form. Does a disordered correlated many-body system serve as its own bath and thermalize? An alternative is that localization in the Fock space~\cite{altshuler1997} prevents the thermalization~\cite{deutsch1991quantum,srednicki1994chaos}. In one-dimensional systems the existence of the many body localization (MBL) transition has been confirmed by extensive numerical studies (see~\cite{abanin2019} for review) and rigorously proven in Ref.~\cite{imbrie2016} for a class of strongly disordered spin chains.

Stability of a localized phase is undermined by the avalanche scenario ~\cite{roeck2017}. Avalanche occurs when a finite ergodic grain, such as an extremal bath or a rear thermal region, delocalize the whole system. Several works~\cite{huse2015,krause2021,brighi2022,seirant2023} studied  small ergodic baths coupled to a localized particles. A discrete localized bath~\cite{hur2025stability} has been found unable to thermalize the localized system. On the other hand, a single mobile impurity~\cite{krause2021,brighi2022,seirant2023} immersed in a localized host system was demonstrated to induce the ergodicity of the whole system.  

An intriguing possibility arises in a strongly correlated system: spin and charge degrees of freedom can serve as independent thermal baths for each other. The question which we address in this work is whether a single charge dopant can destroy the Anderson insulator formed in the spin sector. To that end we consider an $XX$-chain in a random magnetic field, doped with a single hole. We argue that the system displays a peculiar mixture of physics of non-interacting and correlated disordered systems. In the weak disorder limit the hole propagation destabilizes Anderson insulator of spinons and leads to   delocalization in the spin channel. On the other hand, in the strong disorder limit the hole becomes localized and the system as a whole becomes an Anderson insulator.  \cref{fig:fig1} illustrates the model that we consider and the main findings of the work. We contrast our work to the previous studies of disordered one-dimensional $t-J$ models \cite{bonca2017, lemut2017}. In these works, the isotropic spin interactions were considered, which are already delocalized in the weak disorder limit even in the absence of the hole. In the model of our study, the Ising $z-z$ interaction is absent, which guarantees localization of spinons at arbitrarily weak magnetic fields.

The paper is organized as follows. In Sec.~\ref{sec:ModelQA} we present the model and discuss its main properties. We also give qualitative arguments for the expected ergodicity in various basis sets. In Sec.~\ref{sec:Params} we present the order parameters to characterize the expected phase transition. Our numerical results are presented in Sec.~\ref{sec:NumResults}.     Sec.~\ref{sec:Concs} contains our concluding remarks.

\section{Model and qualitative arguments} 
\label{sec:ModelQA}
We study a one-dimensional disordered \(t\)-\(J_{XX}\) chain with a single doped hole. The Hamiltonian is given by
\begin{equation}
H = H_t + H_J + H_h ,
\label{H}
\end{equation}
where
\begin{align}
H_t &= -t \sum_{i=1}^{L}
\sum_{\sigma=\uparrow,\downarrow}
\left(
\tilde c_{i\sigma}^{\dagger}\tilde c_{i+1,\sigma}
+\mathrm{H.c.}
\right), \\
H_J &= J \sum_{i=1}^{L}
\left(
S_i^x S_{i+1}^x + S_i^y S_{i+1}^y
\right), \\
H_D &= \sum_{i=1}^{L} h_i S_i^z .
\end{align}
Here \(\tilde c_{i\sigma}=c_{i\sigma}(1-n_{i\bar\sigma})\) denotes the projected particle annihilation operator enforcing the no-double-occupancy constraint, \(t\) is the hole hopping amplitude, and \(J>0\) is the antiferromagnetic exchange coupling. We impose periodic boundary conditions such that
\begin{equation}
\tilde c_{L+1,\sigma}\equiv \tilde c_{1,\sigma},
\qquad
\mathbf S_{L+1}\equiv \mathbf S_1 .
\end{equation}

The random magnetic fields \(h_i\) are independently drawn from the uniform distribution,
\begin{equation}
h_i \sim \mathcal U(-W,W),
\end{equation}
where \(W\) is the disorder strength. Throughout this work we confine ourselves to the single-hole sector, i.e., the Hilbert space contains exactly one vacant site.

For the frozen hole (\(t=0\)), the spin sector reduces to a set of disordered \(XX\) chains with open boundary conditions (the frozen hole creates a boundary),
\begin{equation}
H_{XX} 
=
J\sum_{i=1}^{L-1}
\left(
S_i^x S_{i+1}^x + S_i^y S_{i+1}^y
\right)
+
\sum_{i=1}^{L-1} h_i S_i^z .
\end{equation}
with the appropriately chosen indexing.
This model can be mapped exactly onto free spinless fermions via the Jordan-Wigner (J-W) transformation~\cite{JWT} , 
\begin{equation}
\begin{aligned}
H_{XX}
={}&
\frac{J}{2}\sum_{i=1}^{L-1}
\left(
\psi_i^\dagger \psi_{i+1}
+\mathrm{H.c.}
\right)
\\
&+
\sum_{i=1}^{L-1}
h_i
\left(
\psi_i^\dagger \psi_i-\frac12
\right),
\end{aligned}
\end{equation}
  where \(\psi_i\) are spinless fermion annihilation operators satisfying canonical anticommutation relations.

At zero-magnetization spin sector (half filling, i.e. $N_f=L/2 $ for J-W fermions), the J-W Hamiltonian describes noninteracting spinless fermions in a random onsite potential, i.e., a one-dimensional Anderson spin insulator for any finite disorder $W>0$. For our model, the single particle spinon wavefunctions are exponentially localized at any energy with the localization length $\xi(E,W)=\frac{6(1-E^2)}{W^2}$ (assuming $J=1$).

The finite hole hopping term \(H_t\) couples the mobile charge degree of freedom to the localized spin/J-W fermion background. The hole motion permutes the spin configuration and thereby induces nontrivial many-body coupling between the charge and spin sectors. In this work we focus on the isotropic point
$J/t=1$, and study how the localization properties of the combined charge-spin system change as a function of disorder strength \(W\).
 \subsection{Clean limit}

In the absence of disorder, \(h_i=0\), translational invariance allows the Hamiltonian to be block diagonalized in sectors of fixed hole momentum. Since there is only a single hole, it is natural to work in the frame in which the hole is at rest. In this representation the hole degrees of freedom can be integrated out, while its momentum appears as a conserved quantum number $p$. This approach was first developed in Ref.~\cite{Sorella} to study the clean limit of the problem.

Let the hole position be denoted as $R$.  In the developed framework one uses the following ansatz:
\begin{equation}
\ket{p,s}
=
\frac{1}{\sqrt L}
\sum_{R=0}^{L-1}
e^{i p R}
T^{R}
\ket{0,s},
\label{ansatz}
\end{equation}
where \(\ket{0,s}\) denotes a state with the hole fixed at a reference site and $\ket{s}$ labels a suitable spin state in $ \ell=L-1$ sites. The operator \(T\) translates  the whole system by one lattice spacing. Then, it can be shown that the state (\ref{ansatz}) is an eigenstate of the Hamiltonian~(\ref{H}), if the suitable state $\ket{s}$ is an eigenstate of the following spin Hamiltonian:
\begin{equation}
    H_p=t\left(e^{ip}T_l+h.c.\right)+J \sum_{i=1}^{l-1}
\left(
S_i^x S_{i+1}^x + S_i^y S_{i+1}^y
\right),
\end{equation}
 where the spins occupy an effective chain of a reduced length $\ell = L-1$, and the operators $T_l$ are translation operators for this squeezed chain. In the J-W representation, one can write the total momentum of the spinons as,
\begin{equation}
P_s
=
\sum_k k\, \psi_k^\dagger \psi_k ,
\end{equation}
where $k$ denotes the momentum basis states with the appropriate quantization for the open boundary conditions. Thus, in the fixed total momentum sector \(p\), the Hamiltonian can be written as
% \begin{equation}
% \hat H_p
% =
% t\exp\left[
% i p
% +
% i\sum_k k\,\psi_k^\dagger \psi_k
% \right]
% +
% \mathrm{H.c.}
% -
% \frac{J}{2}
% \sum_{i=1}^{\ell-1}
% \left(
% \psi_i^\dagger\psi_{i+1}
% +
% \mathrm{H.c.}
% \right).
% \end{equation}
 \begin{equation}
\hat H_p
=
2t\cos(p+P_s)
+
\frac{J}{2}
\sum_{i=1}^{\ell-1}
\left(
\psi_i^\dagger\psi_{i+1}
+
\mathrm{H.c.}
\right).
\end{equation}

The crucial simplification is that the hole hopping depends only on the many-body momentum of the J-W fermions. Therefore the Hamiltonian is diagonal in the many-body momentum basis of the spin chain. Writing the many-body state as
\begin{equation}
\ket{\{k\}}
=
\prod_{j=1}^{N_f}
\psi_{k_j}^\dagger
\ket{0},
\end{equation}
one has
\begin{equation}
P_s\ket{\{k\}}
=
\left(
\sum_{j=1}^{N_f} k_j
\right)
\ket{\{k\}} .
\end{equation}
The corresponding energy eigenvalue is
\begin{equation}
E_p(\{k\})
=
2t\cos\left(
p+\sum_{j=1}^{N_f} k_j
\right)
+
J\sum_{j=1}^{N_f}\cos k_j.
\end{equation}

Thus, in the clean limit, the single-hole problem admits an exact diagonalization in the many-body $k$-basis. Each eigenstate is labeled by the conserved total momentum $p$ and by an occupation pattern $ {n_k} $  of the J-W fermions.  

\subsection{Effect of disorder in momentum space}

We now discuss qualitative effects of a finite random magnetic field in the momentum space. In the laboratory frame the disorder term is
\begin{equation}
H_h=\sum_{i=1, i\neq R}^{L} h_i \psi_j^\dagger \psi_j ,
\end{equation}
with \(h_i\) independently drawn from a uniform distribution. For a hole located at $R$ the Hamiltonian does not depend on the magnetic field $h(R)$. The  disorder potential imposed onto the J-W fermions therefore depends explicitly on the hole coordinate:
\begin{equation}
H_h(R)
=
\sum_{j=1}^{\ell}
h_{R+j}
\left(
\psi_j^\dagger \psi_j-\frac12
\right),
\qquad
\ell=L-1 ,
\end{equation}
where $j$ labels the sites relative to the hole position. Thus, hopping of the hole shifts the entire disorder landscape experienced by the J-W fermions. Unlike the clean case, the spin-chain momentum $P_s$ is no longer conserved, because the random field breaks translation symmetry.

In the momentum space, the disorder term takes the form
\begin{equation}
H_h(R)
=
\frac{1}{\ell}
\sum_{k,q}
\tilde h_{q}
e^{iq R}
\psi_{k+q}^\dagger \psi_k
-
\frac12\sum_j h_{R+j},
\end{equation}
where $ \tilde h_q $  is the Fourier component of the random field. This term scatters the fermions between different momenta and therefore couples different many-body \(k\)-basis configurations, i.e. 
%\begin{equation}
$\ket{\{k \}}
\longleftrightarrow
\ket{\{k '\}}$.
%\end{equation}
Consequently, the exact diagonalization in the many-body momentum basis available in the clean limit is destroyed by the disorder.

Nevertheless, the momentum representation remains useful. It shows that disorder introduces two competing effects. First, for a fixed hole position, the spin sector is equivalent to free  fermions in a random onsite potential and hence is Anderson localized in one dimension. Second, the hole motion alters the disorder configuration of spins. This dynamical rearrangement can hybridize otherwise localized spin states.

At weak disorder, the hole remains sufficiently mobile. Its motion samples many translated disorder landscapes and acts as an internal bath for the spin sector. In the many-body $k$-basis, the disorder-induced scattering matrix elements connect a large number of clean eigenstates whose energy splittings are small compared to the hybridization generated by hole motion. We therefore expect the localized single-particle orbitals of the spin Anderson insulator to become strongly mixed, leading to delocalization of the spin sector.

At strong disorder, by contrast, the random field strongly pins the spin configuration and suppresses coherent hole motion.  The hole hopping changes the local magnetic-field energy of the surrounding spins by an amount of order $W$, producing large energy mismatches between different hole positions. When these mismatches exceed the kinetic scale $t$, resonant hole motion is strongly suppressed. The hole then becomes localized, and the spin sector effectively experiences a static disorder potential. In this regime the ordinary one-dimensional Anderson-localized spin insulator is restored.

\subsection{Hole hopping in the Anderson basis}

A complementary picture of the transition can be obtained in the Anderson basis for the J-W fermions. We will express the Hamiltonian in terms of fermionic operators only, treating the moving hole as a moving boundary subject to quantum fluctuations. To that end let us work in the computational basis for J-W fermions and in open boundary conditions:

\be
|R,\{\bm{n}\}\rangle
=
|R;n_1,n_2,\dots,n_{L-1}\rangle,
\label{eq:JW_basis}
\ee
where $R$ denotes the hole position, and $n_i$ are the occupation numbers for $JW$-fermions. In open boundaries the Hamiltonian for $JW$-fermions reads:
\be
H_{XX} 
= \frac{1}{2} \sum_{i,n} 
\psi_i^{(R)\dagger} \psi_{i+1}^{(R)}+ \mathrm{H.c.}
+ \frac{1}{2} \sum_{i,n} 
h_i^{(R)} \psi_i^{(R)\dagger} \psi_i^{(R)}.
\label{eq:JW occbazis}
\ee

Here an upper index $(R)$ denotes the position of the hole. Note, that we formally put different Fock spaces for the fermions in correspondence to different hole positions. Thus each fermionic operator $\psi_{i+1}^{(R)}$ bears a position of the hole as an index.

Importantly, in open boundary conditions the hole hopping does not change the occupation numbers $\bm{n}$ for the $JW$-fermions: $\tilde{c}^\dagger_{R+1}\tilde{c}_R|R,\{\bm{n}\}\rangle = |R+1,\{\bm{n}\}\rangle$, as illustrated in \cref{fig:fig1}\textbf{(a)} for the dual spin chain. Therefore the hole hopping term can be represented as follows: 
\be
H_t 
= \sum_{n,\{\bm{i}\}} 
\prod_{j \in \{\bm{i}\}} 
\psi_{j}^{(R+1)\dagger} 
\psi_{j}^{(R)} 
+ \mathrm{h.c.}.
\label{eq:hole hopping occbazis}
\ee
Here $\{\bm{i}\}$ denotes an arbitrary set of possible fermionic indices. The disordered potential is different for each hole position $(R)$ -- the energy of the system does not depend on the magnetic field at the site $R$. Therefore, strictly speaking, the $XX$-Hamiltonian has to be diagonalized in the Anderson basis individually for each $R$ : 

\be
\begin{split}
    H_{XX} 
&= \sum_\alpha \varepsilon^{(R)}_\alpha \, 
\phi_\alpha^{(R)\dagger} \phi_\alpha^{(R)} \\
\phi_\alpha^{(R)} 
&= \sum_i U_{\alpha i}^{(R)} \psi_i^{(R)}.
\end{split}
\label{eq:JW Andbazis}
\ee

Now let us note that in the strong disorder limit most of the interacting terms in \cref{{eq:hole hopping occbazis}} act trivially in the Anderson basis, changing merely the index $(R)$. Let us consider for simplicity the single-particle (single spinon) sector. Consider the matrix elements of the hole hopping Hamiltonian in the Anderson basis: 

\be
\begin{split}
\langle 0 | 
\phi_{\alpha}^{(R+1)} 
\hat{H}_t 
\phi_{\beta}^{(R)\dagger} 
| 0 \rangle
&=
\left\langle 0 \left| 
\sum_{ij}
U_{\alpha i}^{(R+1)}
\psi_i^{(R+1)}
\hat{H}
U_{j\beta}^{\dagger(R)}
\psi_j^{(R)\dagger}
\right| 0 \right\rangle\\
&=
\sum_i
U_{\alpha i}^{(R+1)}
U_{i\beta}^{\dagger (R)}.
\end{split}
\label{eq: hole induced hopping}
\ee

Suppose that the centers of localization $\bm{r}_\alpha$ and $\bm{r}_\beta$ are well separated from the hole: $|\bm{r}_{\alpha/\beta}-R| >> \xi$. Then one can neglect the difference between the matrices $U_{\alpha i}^{(R+1)}$ and $U_{i\beta}^{(R)}$, since they are exponentially close to each other: 

\be
\sum_i 
U_{\alpha i}^{(R+1)}
U_{i\beta}^{\dagger (R)}
= \delta_{\alpha\beta} + O(e^{|\bm{r}_\alpha - R|/\xi}),
\label{eq: different Us}
\ee
where we assumed without the loss of generality that $|\bm{r}_\alpha - R| < |\bm{r}_\beta- R|$. Equation \cref{eq: different Us} formalizes an intuitively clear idea: the hole hopping can induce scattering between the Anderson orbitals only in the vicinity of the hole. 

The same logic applies to the sectors with many J-W fermions. For the half-filled system that we consider, the hole hopping term \cref{eq:hole hopping occbazis} formally is an $(L-1)/2$-particle operator. However, \cref{eq: different Us} dictates that it effectively acts as a $\lceil \xi \rceil$-particle operator, where $\lceil \xi \rceil$ denotes the ceiling function. In the limit of weak disorder, the hole hopping induces non-trivial many-body interaction between the Anderson orbitals leading as we shall see to thermalization. In the strong disorder limit, where the localization length becomes of the order of a single site spacing the system becomes effectively gaussian and therefore the Anderson localization restores.  

 \section{Characterization of the localization transition}
\label{sec:Params}

To characterize the localization properties of the disordered single-hole problem, we analyze spectral statistics and eigenstate structure of the full charge-spin Hamiltonian. In particular, we employ the same diagnostics used in studies of MBL~\cite{Pino,Luitz,Oganesyan} adapted here to the single-hole sector.

An important quantity characterizing the eigenstates of the disordered Hamiltonian is the ratio of consecutive level spacings,
\begin{equation}
r_i=
\frac{\min(\Delta_i,\Delta_{i+1})}
{\max(\Delta_i,\Delta_{i+1})},
\qquad
\Delta_i=\epsilon_i-\epsilon_{i-1},
\end{equation}
where \(\epsilon_i\) are the ordered many-body energy eigenvalues for a given disorder realization. In the delocalized (chaotic) regime the level-spacing statistics follows the Wigner--Dyson distribution of the Gaussian orthogonal ensemble (GOE), whereas in the localized regime no level repulsion is expected and the statistics becomes Poissonian. The corresponding disorder-averaged values are 
\begin{equation}
\langle r\rangle_{\rm P}=2\ln 2-1 \approx 0.386,
\qquad
\langle r\rangle_{\rm WD} \approx 0.53.
\end{equation}
A crossover of \(\langle r\rangle\) from the Wigner--Dyson to Poisson value therefore signals the onset of localization.

\vspace{0.3cm}

We further characterize eigenstate localization in Hilbert space through participation entropies and fractal dimensions. In the single-hole sector of a chain with \(L\) sites and fixed total magnetization \(S^z_{\rm tot}\), the Hilbert-space dimension is
\begin{equation}
{\cal N_H}
=
L\binom{L-1}{N_\uparrow},
\end{equation}
where \(N_\uparrow\) is the number of up spins among the \(L-1\) occupied sites. We work in the computational basis
\begin{equation}
|R,\lbrace\sigma\rbrace\rangle
=
|R;\sigma_1,\sigma_2,\dots,\sigma_{L-1}\rangle ,
\end{equation}
where \(R\) denotes the hole position and \(\sigma_i=\uparrow,\downarrow\) specifies the spin configuration of the occupied sites.

For the many-body eigenstate \(|\alpha\rangle\), with wavefunction amplitudes $\psi_\alpha(s)=\langle s|\alpha\rangle$, the participation entropies are defined as
\begin{equation}
S_q=
\frac{1}{1-q}
\ln\left(
\sum_{s=1}^{{\cal N_H}}
|\psi_\alpha(s)|^{2q}
\right).
\end{equation}
Their scaling with Hilbert-space dimension defines the fractal dimensions \(D_q\),
\begin{equation}
S_q
\xrightarrow{{\cal N_H}\rightarrow\infty}
D_q\ln({\cal N_H}).
\end{equation}
Localized eigenstates occupying only a finite number of basis states satisfy \(D_q=0\), whereas fully ergodic states have \(D_q=1\). Intermediate values \(0<D_q<1\) correspond to nonergodic extended (multifractal) states. In the following we focus on the Shannon entropy limit \(q\to1\),
\begin{equation}
S_1=
-\sum_s |\psi_\alpha(s)|^2 \ln |\psi_\alpha(s)|^2 ,
\end{equation}
with the corresponding disorder-averaged fractal dimension $D_1=D.$
 
To directly characterize localization of the doped hole in real space, we analyze the localization properties of the hole probability distribution associated with each many-body eigenstate. While the participation entropies introduced above probe the structure of the full many-body wavefunction, they do not themselves distinguish whether the localization arises primarily from the spin sector or from the charge sector. To isolate the charge degree of freedom, we consider the reduced hole density profile obtained by tracing over the spin configuration.

For the many-body eigenstate \(|\alpha\rangle\), the probability of finding the hole at site \(j\) is
\begin{equation}
\rho_\alpha(j)
=
\sum_{\{\sigma\}}
\left|
\langle j,\{\sigma\}|\alpha\rangle
\right|^2 ,
\label{rhoJ}
\end{equation}
where \(|j,\{\sigma\}\rangle\) denotes a basis state with the hole fixed at site \(j\) and spin configuration \(\{\sigma\}\) on the remaining \(L-1\) sites. By construction,
\begin{equation}
\sum_{j=1}^{L}\rho_\alpha(j)=1.
\end{equation}

We define the generalized inverse participation ratios of the hole distribution,
\begin{equation}
P_q^{(h)}
=
\sum_{j=1}^{L}
\rho_\alpha(j)^q ,
\end{equation}
and the corresponding disorder-averaged hole participation entropies  $S_q^{(h)}$.
 
Their scaling with system size defines the fractal dimensions of the hole wavefunction, $D_q^{(h)}$. We further also restrict outselves with the Shannon limit and calculate only $D_1^{(h)}=D_h.$
 
The hole fractal dimension provides a direct measure of charge localization and complements the many-body diagnostics discussed above. In the weak-disorder regime, where the hole remains mobile and delocalizes the spin background, we expect \(D_h\approx 1\), reflecting an extended hole wavefunction. In the strong-disorder regime, where the hole becomes pinned by the disordered spin environment, \(D_h\) decreases towards zero, signaling localization of the charge degree of freedom. Calculation of \(D_h\) alongside with the many-body fractal dimensions $D$ therefore allows us to characterize localization of the hole and localization in the full many-body Hilbert space.

As an additional probe of eigenstate hybridization, we evaluate the Kullback--Leibler divergence between neighboring eigenstates in energy~\cite{Pino},
\begin{equation}
KL=
\sum_{s=1}^{{\cal N_H}}
|\psi_\alpha(s)|^2
\ln
\left(
\frac{|\psi_\alpha(s)|^2}{|\psi_{\alpha+1}(s)|^2}
\right).
\end{equation}
In the localized regime neighboring eigenstates occupy largely distinct regions of Hilbert space, implying exponentially small overlap and hence \(KL\to\infty\) in the thermodynamic limit. In contrast, in the delocalized regime neighboring eigenstates strongly hybridize and are supported on similar sets of basis states, yielding finite \(KL\). A sharp crossover in \(KL\) therefore provides an additional indicator of the localization transition.

The level statistics, fractal dimensions, and Kullback--Leibler divergence provide complementary diagnostics of the disorder-driven transition between the weak-disorder regime, where the mobile hole delocalizes the spin background, and the strong-disorder regime, where localization of the hole restores the spin Anderson insulator.

We focus on eigenstates in the middle of the many-body spectrum (we also show the full spectrum properties only for $L=13$), where hybridization effects are the strongest and finite-size effects are minimized. Exact diagonalization is performed in fixed-magnetization symmetry sector with $S^z_{tot}=0$. For system sizes \(L=\{ 9,11,13,17 \}\), we obtain eigenstates near the spectral center using shift-invert exact diagonalization~\cite{Pietracaprina}. The number of disorder realizations ranges from \(N_{\rm dis}=10^4\) for the smallest systems to $10^2$  for the largest sizes. All observables are first averaged over the selected eigenstates within each realization and subsequently over disorder realizations.
 
In addition to eigenstate diagnostics, we characterize the localization properties of the hole through the real-time growth of the width of the initially localized hole. To this end, we employ the time-evolving block decimation algorithm (TEBD)~\cite{Paeckel} to consider the time-evolution of an initial product state:
\begin{equation}
|\psi_0\rangle
=
|(L-1)/2,\mathrm{N\acute eel}\rangle ,
\end{equation}
corresponding to the hole initially placed at the center of the chain and the spins prepared in a N\'eel configuration in the remaining sites. The number of disorder realizations used to average time-dependent observables were $N_{dis}\sim 10^2$. To guarantee the accurate calculation of the dynamics of the many-body state, we always used $\delta t=0.01$ and kept the singular values $s<10^{-9}$ in the TEBD algorithm. 

For each time-evolved state \(|\psi(t)\rangle\), we compute the hole distribution $\rho(j,t)$ via Eq.~(\ref{rhoJ})  and the mean square deviation of the hole distribution,
\begin{equation}
 \sigma^2(t)=\sum_jj^2\rho(j,t)-\left(\sum_jj\rho(j,t)\right)^2.
\end{equation}

The temporal growth of $\sigma^2(t)$ provides a dynamical marker distinguishing the Anderson-localized from delocalized behavior~\cite{lemut2017}. In the Anderson-localized regime, where the many-body dynamics is effectively reduced to that of noninteracting localized quasiparticles, the initial packet spreads only over a finite localization length and therefore $\sigma^2(t)$ exhibits saturation. On the other hand, for the delocalized phase one expects diffusion of the initially localized hole packet, i.e. $\sigma^2(t)\propto t$.
 
\section{Numerical results}
\label{sec:NumResults}
\begin{figure}[t]
\centering
\includegraphics[width=0.9\columnwidth]{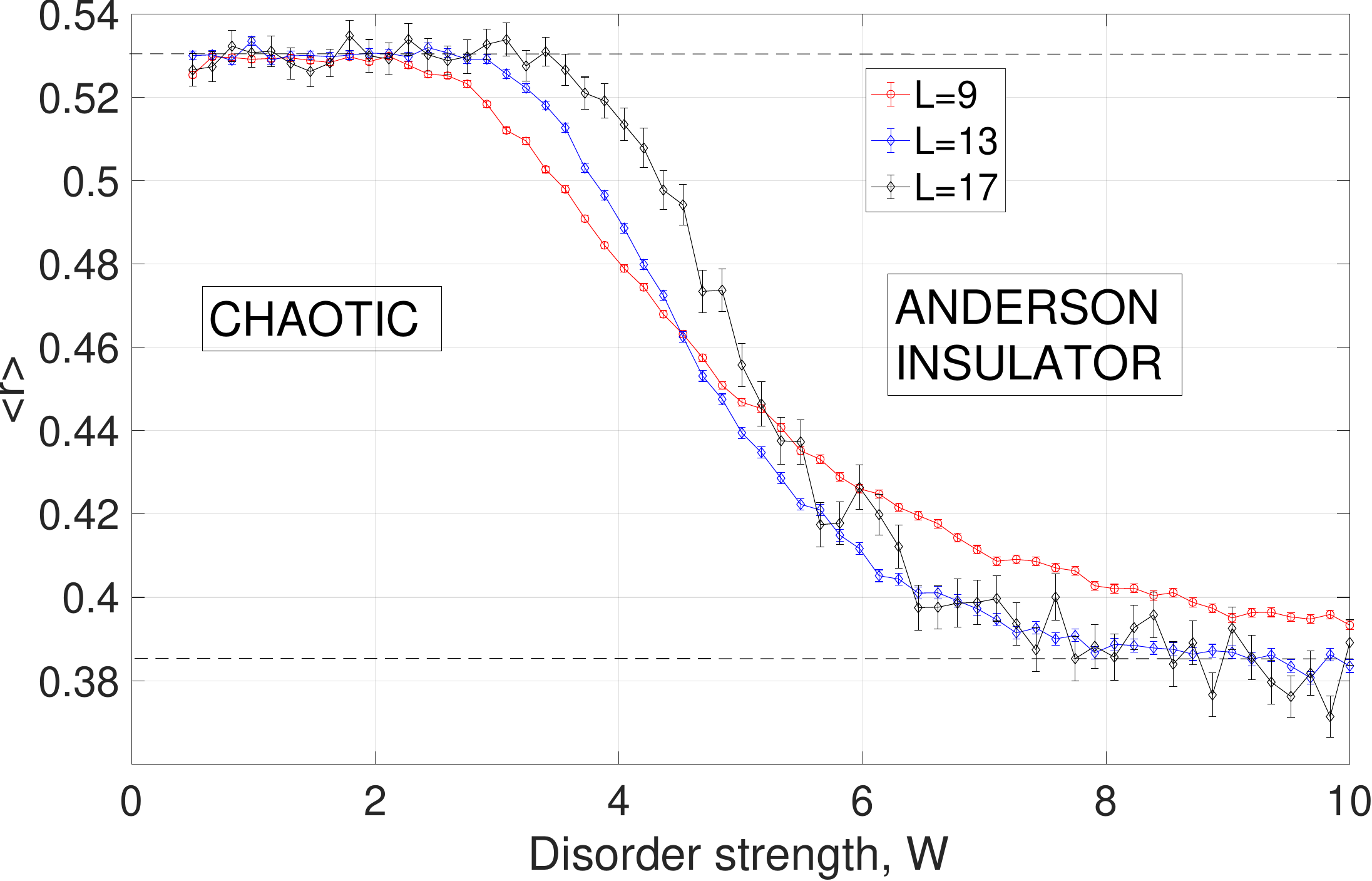}
\caption{
Disorder dependence of the disorder-averaged adjacent-gap ratio \(\langle r\rangle\) for the single-hole doped disordered \(XX\) chain at \(t/J=1\), shown for system sizes \(L=\lbrace9 ,13,17\rbrace\). At weak disorder the level statistics is close to the Wigner--Dyson value \(\langle r\rangle_{\rm WD}\approx0.53\), indicating strong level repulsion and delocalized many-body eigenstates. With increasing disorder, \(\langle r\rangle\) decreases continuously towards the Poisson value \(\langle r\rangle_{\rm P}\approx0.386\), signaling the suppression of level repulsion and onset of localization. The crossover sharpens with increasing system size, consistent with a disorder-driven localization transition of the coupled hole-spin system.
}
\label{fig:rstat}
\end{figure}

  We begin with examining the spectral statistics of the disordered single-hole Hamiltonian through the adjacent-gap ratio \(\langle r\rangle\), shown in Fig.~\ref{fig:rstat}. In the weak-disorder regime, \(\langle r\rangle\) is close to the Gaussian orthogonal ensemble value \(\langle r\rangle_{\rm WD}\approx0.53\), indicating pronounced level repulsion and thus nonintegrable, delocalized many-body eigenstates. 
 \begin{figure}[ht]
\centering
\includegraphics[width= \columnwidth]{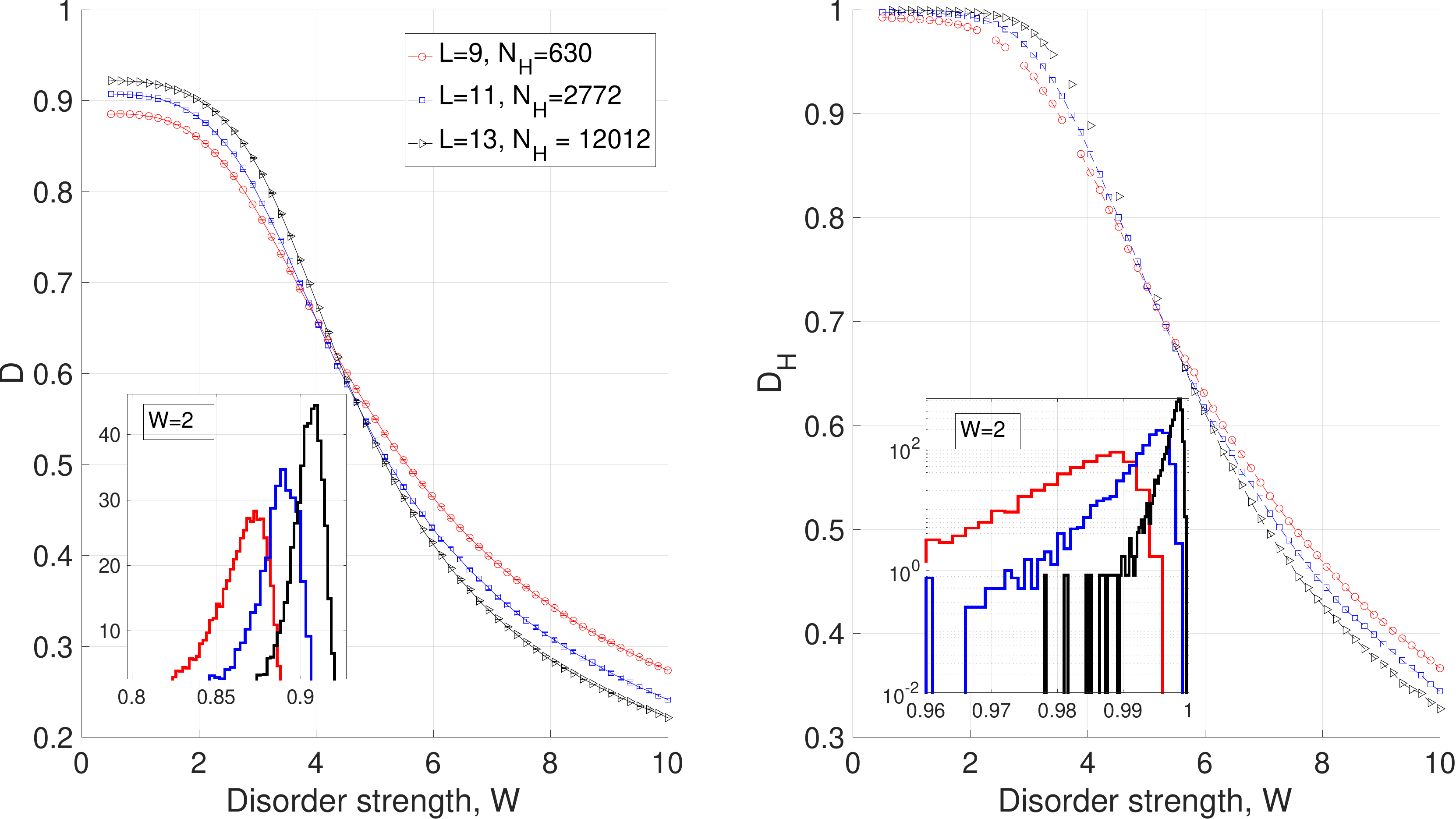}
\caption{
Disorder dependence of the Shannon fractal dimensions obtained from the full many-body eigenstates, \(D\) (left panel), and from the hole wavefunctions, \(D_H\), (right panel) for system sizes \(L=9,11,13\). The many-body fractal dimension \(D\) is extracted from the participation entropy of the full eigenstate in the computational basis, while \(D_H\) is obtained from the reduced hole probability distribution in real space. Insets: histograms of \(D\) (left) and \(D_H\) (right) at weak disorder \(W=2\). Errorbars are smaller than the symbol sizes.
}
\label{fig:fractal}
\end{figure}

As the disorder strength increases, \(\langle r\rangle\) decreases monotonically and approaches the Poisson limit \(\langle r\rangle_{\rm P}\approx0.386\), characteristic of localized spectra without level repulsion. This demonstrates that sufficiently strong disorder suppresses the hole-induced hybridization between many-body configurations and drives the system into the localized regime.
 
The crossover between the two limits occurs around \(W/t\sim 5\), where the finite-size curves exhibit the strongest drift and maximum slope. Moreover, the crossover becomes sharper with increasing the system size, suggesting that this regime corresponds to the finite-size precursor of localization transition in the thermodynamic limit. Taken together, the spectral statistics provides the first clear evidence for two distinct regimes in the disordered single-hole problem: a weak-disorder ergodic phase in which the hole delocalizes the spin Anderson insulator, and a strong-disorder localized phase in which the hole itself localizes and the spin insulator is restored.
 
We next analyze the fractal properties of the eigenstates. Figure~\ref{fig:fractal} shows the disorder dependence of the Shannon fractal dimension of the full many-body eigenstates, \(D\), together with the fractal dimension of the hole wavefunction, \(D_H\).

Several important features are immediately apparent. First, throughout the weak-disorder regime the hole remains almost fully extended, with \(D_H\approx1\), while the many-body fractal dimension is significantly smaller, \(D<1\). This demonstrates that even when the charge degree of freedom is delocalized across the entire chain, the full many-body wavefunction is not fully ergodic in Hilbert space for these system sizes. Instead, the eigenstates occupy only a fractal subset of the available basis states, reflecting residual structure inherited from the localized spin background.
\begin{figure}[t]
\centering
\includegraphics[width=0.8\columnwidth]{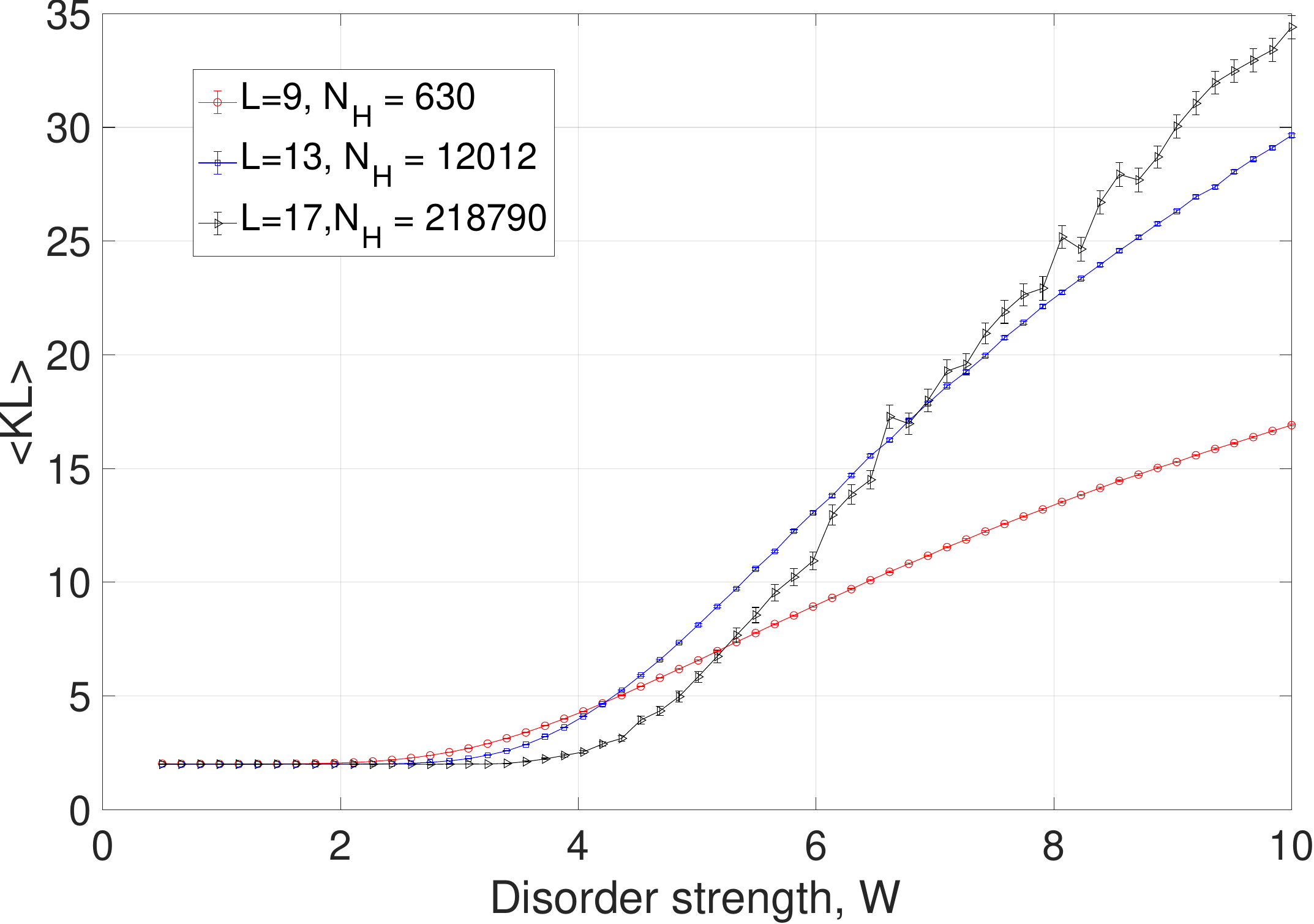}
\caption{
Disorder dependence of the disorder-averaged Kullback--Leibler divergence \(\langle KL\rangle\) between neighboring many-body eigenstates for system sizes \(L=9,13,17\).  
}
\label{fig:KL}
\end{figure}
Second, both \(D\) and \(D_H\) decrease monotonically with increasing disorder, indicating gradual localization of both the many-body eigenstates and the hole itself. However, the onset of this decrease occurs at different (although close) disorder strengths. The many-body fractal dimension \(D\) begins to decrease already at relatively weak disorder, whereas \(D_H\) remains pinned very close to unity up to substantially larger disorder. However, we expect that in the thermodynamic limit there is a single critical point.

The insets of Fig.~\ref{fig:fractal} show the distributions of \(D\) and \(D_H\) at weak disorder \(W=2\). The distributions become narrower with increasing thesystem size and their peaks drift systematically, indicating that both quantities are self-averaging and well defined in the large-$L$ limit. In particular, the histogram of \(D_H\) is sharply concentrated near unity, confirming that the hole is essentially fully delocalized in this regime, while the broader distribution of \(D\) is centered below unity. Nevertheless, the mean value of \(D\) increases with the system size and for $L\rightarrow\infty$ one expects $D\rightarrow1$.

At stronger disorder, both fractal dimensions decrease significantly, with \(D_H\) eventually departing from unity and approaching substantially lower values. This marks the onset of localization of the hole itself and coincides with the crossover towards Poisson level statistics observed in Fig.~\ref{fig:rstat}. The simultaneous suppression of \(D\) and \(D_H\) therefore supports the interpretation that strong disorder localizes the hole and restores the localized character of the spin background.
\begin{figure*}[t]
\centering
\includegraphics[width=0.8\textwidth]{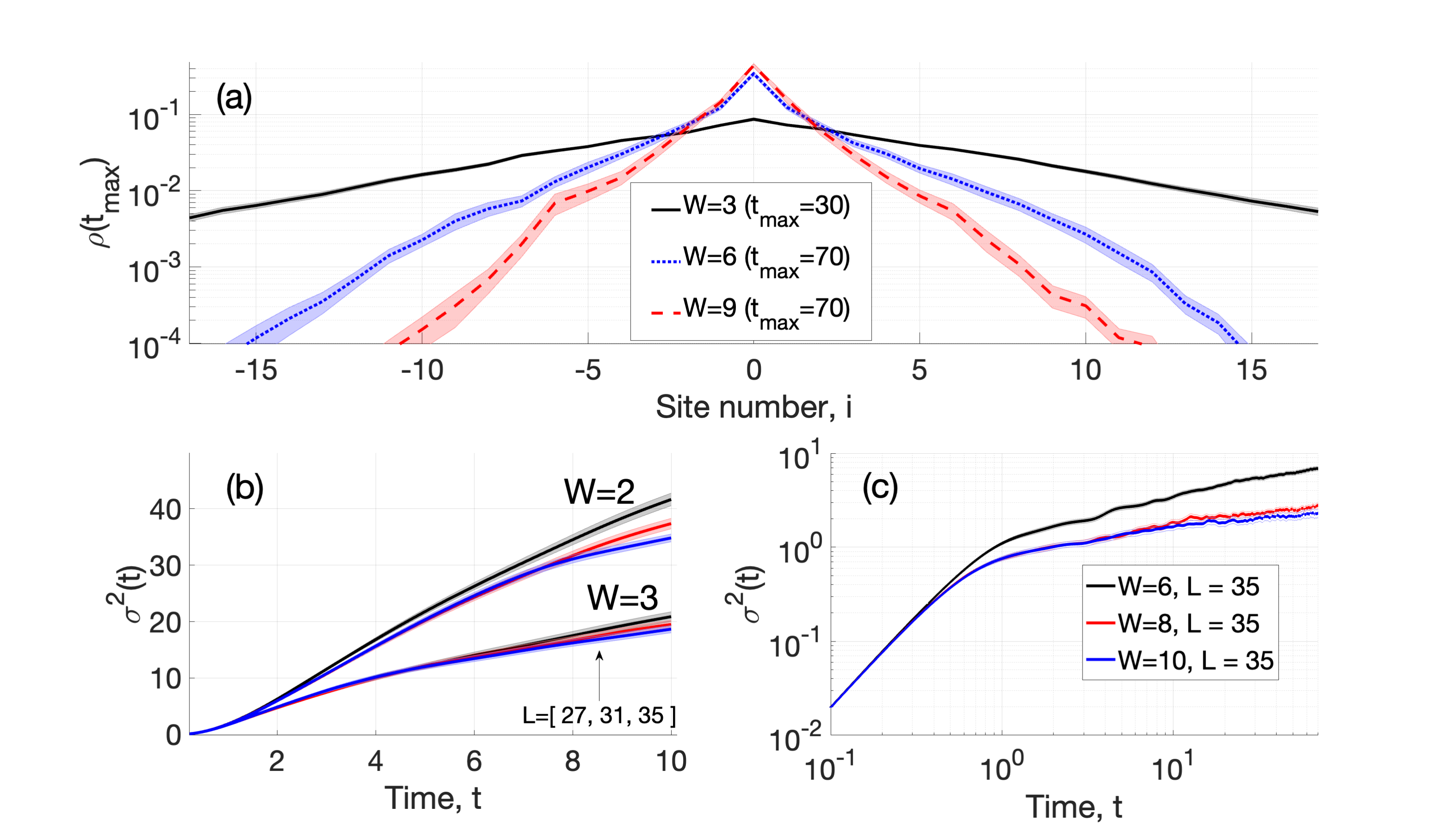}
 \caption{
Real-time dynamics of the doped hole. 
(a) Hole probability distribution \(\rho(j,t_{\rm max})\) at the largest accessible time for disorder strengths \(W=3,6,\) and \(9\). At weak disorder the hole distribution is broad, whereas increasing \(W\) produces a progressively sharper, exponentially localized profile around the initial hole position. Shaded regions indicate statistical error.
(b) Mean-square displacement \(\sigma^2(t)\) for weak disorder \(W=2\) and \(W=3\) and different system sizes \(L=27,31,35\).   
(c) Mean-square displacement at stronger disorder \(W=6,8,10\) for \(L=35\), shown in logarithmic scales. The progressive slowing down and tendency towards saturation demonstrate the emergence of a finite localization length of the hole with increasing disorder.
}
\label{fig:hole_dynamics}
\end{figure*}
As a further probe of eigenstate hybridization in Hilbert space, we examine the Kullback--Leibler divergence between neighboring many-body eigenstates, shown in Fig.~\ref{fig:KL}. This quantity directly measures the similarity of adjacent eigenstates in the computational basis and therefore provides a sensitive diagnostics of localization.

In the weak-disorder regime, \(\langle KL\rangle=2\) remains constant, and exhibits a system-size dependence. This indicates that neighboring eigenstates strongly overlap in Hilbert space and are composed of largely the same basis configurations, as expected when disorder-induced localized configurations are efficiently hybridized by the motion of the hole. The finite and nearly size-independent value of \(\langle KL\rangle\) is therefore consistent with delocalized and chaotic many-body eigenstates.

As disorder increases, \(\langle KL\rangle\) begins to grow rapidly and develops strong system-size dependence. This reflects a progressive reduction of overlap between neighboring eigenstates. Adjacent eigenstates become supported in distinct regions of Hilbert space, indicating that hybridization between many-body configurations is suppressed. In the strong-disorder regime, \(\langle KL\rangle\) grows approximately monotonically with the system size, consistent with the expectation that \(KL\) diverges in the thermodynamic limit for localized eigenstates.
 
The crossover in \(\langle KL\rangle\) occurs in the same disorder window identified by the level-statistics and fractal-dimension diagnostics, \(W/t\sim 5\). This agreement provides further evidence that the system undergoes a disorder-driven localization transition separating a weak-disorder regime from a strong-disorder regime in which the hole localizes and neighboring many-body eigenstates become effectively decoupled in Hilbert space.
 
 To further clarify the nature of the strongly disordered regime, we examine the real-time dynamics of the hole distribution $\rho(j,t)$  following a global quench from the product state \(|(L-1)/2,\mathrm{N\acute eel}\rangle\).   Figure~\ref{fig:hole_dynamics}(a) shows the hole density profile \(\rho(j,t_{\rm max})\) at the largest accessible simulation times up to $t_{max}=70$ (in units of $[1/t]$) for several disorder strengths at fixed $L=35$. A pronounced qualitative change is observed upon increasing disorder strength. For weak disroder \(W=3\), the distribution at $t_{max}=30$ is already broad and extends over the entire accessible sites, consistent with a delocalized hole. In contrast, for \(W=6\) and, even more clearly, for \(W=9\), the density develops an exponentially decaying profile, $\rho(j,t_{max})\sim e^{-|j-j_0|/\xi_h}$, away from its initial position. The hole localization length $\xi_h$ decreases strongly with increasing \(W\), providing direct evidence for disorder-induced localization of the hole.

This conclusion is supported by the time dependence of the mean-square displacement $\sigma^2(t)$ at short time scale $t_{max}\sim 10$. At weak disorder, shown in Fig.~\ref{fig:hole_dynamics}(b), \(\sigma^2(t)\)  for \(W=2\) and \(W=3\) the packet exhibits diffusive behavior $\sigma^2(t)\propto t$. The hole therefore remains mobile in precisely the regime where the level statistics exhibits strong level repulsion and the hole fractal dimension is close to unity.

The situation is qualitatively different at strong disorder [Fig.~\ref{fig:hole_dynamics}(c)] (considered up to $t_{max}\sim10^2$). For \(W=6,8,\) and \(10\), the initial spreading is followed by a pronounced slowing down and a tendency of \(\sigma^2(t)\) towards a finite value. Moreover, increasing the disorder systematically reduces the long-time spatial extent of the hole. Such saturation of the mean-square displacement is the dynamical signature of a finite localization length and is consistent with the exponentially localized density profiles in Fig.~\ref{fig:hole_dynamics}(a).

The real-time dynamics therefore independently corroborates the physical picture obtained from the spectral and eigenstate diagnostics. At weak disorder the hole spreads through the system and provides a dynamical channel capable of hybridizing the Anderson-localized spin configurations. Upon increasing \(W\), the hole acquires a finite localization length.  

\begin{figure}[t]
\centering
\includegraphics[width=0.9\columnwidth]{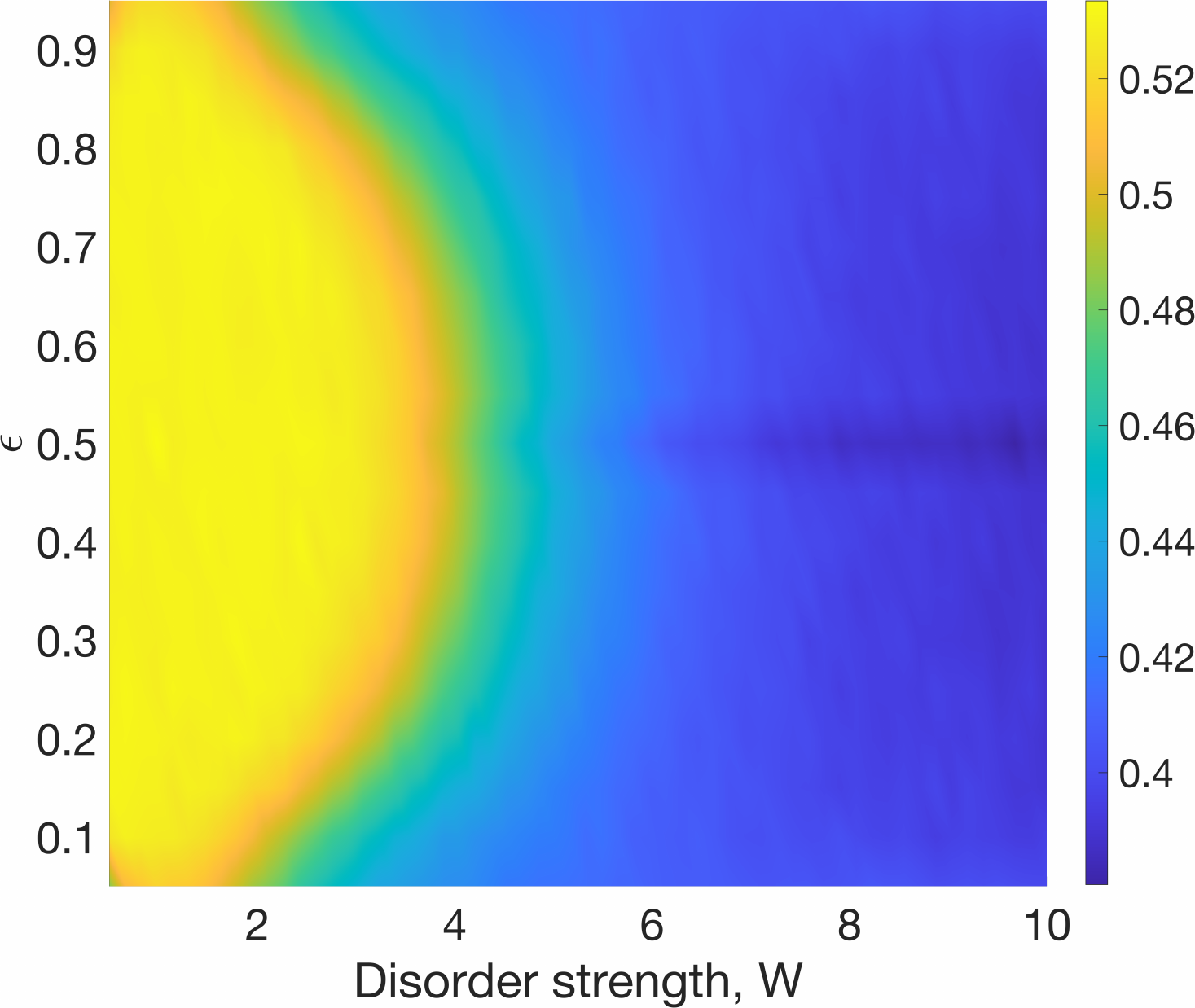}
\caption{
Energy-resolved adjacent-gap ratio \(\langle r\rangle\) as a function of disorder strength \(W\) and reduced energy density \(\epsilon\), where \(\epsilon=0\) corresponds to the ground state and \(\epsilon=1\) to the highest excited state. The color scale represents the disorder-averaged value of \(\langle r\rangle\), interpolating between Wigner--Dyson statistics (yellow, \(\langle r\rangle\approx0.53\)) and Poisson statistics (blue, \(\langle r\rangle\approx0.386\)). The crossover from delocalized to localized spectral statistics occurs over the entire spectrum, with only weak dependence on energy density, indicating the absence of a pronounced mobility edge in the studied parameter regime.
}
\label{fig:ern}
\end{figure}

To investigate whether localization sets in uniformly across the spectrum or exhibits energy dependence, we compute the adjacent-gap ratio resolved in normalized energy density. Figure~\ref{fig:ern} shows the disorder-averaged value of \(\langle r\rangle\) as a function of disorder strength \(W\) and reduced energy density \(\epsilon\), where \(\epsilon=0\) corresponds to the ground state and \(\epsilon=1\) to the highest excited state, i.e. the energy is given by $E=(E_{max}-E_{min})\epsilon+E_{min}.$

At weak disorder, the entire spectrum is characterized by values of \(\langle r\rangle\) close to the Wigner--Dyson limit, indicating chaotic behavior. As disorder increases, \(\langle r\rangle\) decreases continuously towards the Poisson value across all energy densities, demonstrating that localization develops globally rather than being restricted to a particular spectral window.

Importantly, the crossover line separating the delocalized and localized regimes is only weakly dependent on energy density. Although a mild bending of the crossover boundary is visible near the spectral edges, no pronounced mobility edge is observed within the accessible system sizes and disorder range. 

The broad vertical crossover region centered around \(W/t\sim5\) is consistent with the disorder scale extracted from the energy-averaged level statistics discussed above. The absence of strong energy dependence further supports the interpretation that the localization transition is governed primarily by the competition between the hole mobility and disorder strength, rather than by the specific energy density of the many-body eigenstates.
\section{Conclusions}
We considered destabilization of one-dimensional spin Anderson insulator by a mobile doped hole within the $t-J_{XX}$ model. Using exact diagonalization method, we demonstrated that when the disorder is weak, the charge carrier destroys localization of the spin channel. This is demonstrated by the exhibited level repulsion and delocalized many-body states of the system. On the other hand, at strong disorder the mobile carrier localizes with field-dependent localization length and the Anderson insulator of spins is restored. We provide qualitative arguments for the transition in various basis sets. The presented model serves as a toy model for further analytical investigation of destabilization mechanism of Anderson spin insulator coupled to charge bath.
 \label{sec:Concs}

\begin{acknowledgements}
The authors thank B. L. Altshuler and V. E. Kravtsov for useful discussions. This work was financially supported by the Russian Science Foundation (project No. 25-72-00135). This research was also supported in part
through computational resources of the HPC facilities at
HSE University. A.A.M. acknowledges the support from European Research Council (LATIS project).

\end{acknowledgements}

\end{document}